\documentclass{article}

\usepackage[english]{babel}
\usepackage{cite}
\usepackage{leftidx}
\usepackage[letterpaper,top=2cm,bottom=2cm,left=3cm,right=3cm,marginparwidth=1.75cm]{geometry}

\usepackage[affil-it]{authblk}
\usepackage{amsmath}
\usepackage{amsfonts}
\usepackage{subfigure}
\usepackage{graphicx}
\usepackage[colorlinks=true, allcolors=blue]{hyperref}
\usepackage{cleveref}
\usepackage[page]{appendix}
\usepackage[font=small,labelfont=bf]{caption}

\DeclareMathOperator*{\argmax}{arg\,max}
\font\myfont=cmr12 at 23pt
\title{\myfont Social media battle for attention: \\opinion dynamics on competing networks}
\author[1,2]{Andrea Somazzi\thanks{{\href{mailto:andrea.somazzi@imtlucca.it}{andrea.somazzi@imtlucca.it}}}}
\author[3]{Giuseppe M. Ferro\thanks{{\href{mailto:giuseppe.ferro@princeton.edu}{giuseppe.ferro@princeton.edu}}}}
\author[1,4,5]{Diego Garlaschelli}
\author[3]{Simon A. Levin}

\affil[1]{\footnotesize IMT School for Advanced Studies, Piazza S. Francesco 19, 55100 Lucca, Italy}
\affil[2]{Scuola Normale Superiore, Piazza dei Cavalieri 7, 56126 Pisa, Italy}
\affil[3]{Department of Ecology and Evolutionary Biology, Princeton University, Princeton, NJ 08544}
\affil[4]{Lorentz Institute for Theoretical Physics, Niels Bohrweg 2, 2333 CA Leiden, The Netherlands}
\affil[5]{INdAM-GNAMPA Istituto Nazionale di Alta Matematica, Italy}
\date{}
\begin{document}
\maketitle
\vspace{-1.2cm}
\begin{center}
    \footnotesize \textbf{Author contribution}
\end{center}
\vspace{-0.3cm}
\footnotesize A.S. and G.M.F. conceived the idea and design, developed the mathematical framework, performed all simulations, analyze and interpret the results and wrote the manuscript. \\D.G and S.A.L. provided critical discussions, guidance, and supervision throughout the research process.
\vspace{1cm}
\begin{abstract}
In the age of information abundance, attention is a coveted resource. Social media platforms vigorously compete for users' engagement, influencing the evolution of their opinions on a variety of topics. With recommendation algorithms often accused of creating ``filter bubbles'', where like-minded individuals interact predominantly with one another, it's crucial to understand the consequences of this unregulated attention market. To address this, we present a model of opinion dynamics on a multiplex network. Each layer of the network represents a distinct social media platform, each with its unique characteristics. Users, as nodes in this network, share their opinions across platforms and decide how much time to allocate in each platform depending on its perceived quality. Our model reveals two key findings. i) When examining two platforms --- one with a neutral recommendation algorithm and another with a homophily-based algorithm --- we uncover that even if users spend the majority of their time on the neutral platform, opinion polarization can persist. ii) By allowing users to dynamically allocate their social energy across platforms in accordance to their homophilic preferences, a further segregation of individuals emerges. While network fragmentation is usually associated with ``echo chambers'', the emergent multi-platform segregation leads to an increase in users' satisfaction without the undesired increase in polarization. These results underscore the significance of acknowledging how individuals gather information from a multitude of sources. Furthermore, they emphasize that policy interventions on a single social media platform may yield limited impact.
\end{abstract}

\section{Significance statement}
Understanding how people consume information on social media is key to unveil its influence on political opinions and shaping effective platform regulation policies. A Facebook study suggests that their recommendation algorithm, suggesting like-minded content, has a limited impact on political attitudes, as such beliefs remain unchanged even when individuals are exposed to more diverse content. Our opinion dynamics model challenges this study, demonstrating that the interplay between diverse and homophilic content recommendations across various platforms can sustain opinion polarization, highlighting the importance of multi-media information gathering. Our model also reproduces the observed relationship between opinions' stance and social media preferences. These insights underscore the intricate relationship between recommendation algorithms, news consumption, and opinion dynamics in the digital age, with the potential to inform and guide policy decision-making within the ever-changing landscape of information dissemination.
\section{Introduction} \label{sec:intro}
In our contemporary landscape, online social networks have evolved into pivotal platforms for the acquisition of political news \cite{bond201261,shearer2019americans}. These modern communication platforms have several advantages for democracy: they simplify access to information, boost citizen participation, enable individuals to express their views, counteract misinformation, and enhance transparency and responsibility in political actions. Ideally, individuals can tap into social media to encounter a range of ideological perspectives and consequently make more informed choices \cite{barbera2015tweeting, sunstein2018social,guess2018avoiding}. An extensive literature, empirical \cite{bakshy2015exposure,del2016spreading,garimella2018political,huszar2022algorithmic,levy2021social,nyhan2023like} and theoretical \cite{holme2006nonequilibrium,wang2020public,santos2021link,baumann2020modeling}, pertains to how social media influence opinion dynamics, fostering (or not) the appearance of ``echo-chambers'' where individuals are mainly connected with like-minded peers. Echo chambers are often believed to stem from the recommendation systems utilized by social media platforms, which tend to link individuals with similar views. This phenomenon contributes to the rise of opinion polarization \cite{garimella2018political}. Furthermore, these automated algorithms interact with the cognitive limitations of individuals, who tend to gravitate towards information that aligns with their existing beliefs and actively avoid contradictory information \cite{hart2009feeling}. It is thus crucial to distinguish between the influence of algorithms and inherent human tendencies when examining the genesis of echo chambers. A recent empirical study \cite{nyhan2023like} has shown how increasing the amount of cross-cutting content on Facebook does not significantly alter political opinions, thereby suggesting that social media recommendation algorithms may not contribute to opinion polarization. They note however that political information on Facebook is mainly incidental ($6.7\%$ of the total news consumption), and thus people might get political information from other sources. More in general, the above cited studies focus on the effect of only one social media platform, while contemporary news consumption is characterized by its reliance on a multitude of sources \cite{hasebrink2006media}. People exhibit different news repertoires depending on a variety of needs \cite{zillmann2000mood, rubin2002uses,tandoc2019platform}. Several scholars \cite{holbert2009theory,yuan2011news,hasebrink2012media} stress that an adequate study of political opinions must consider the interplay between news repertoires and political communication processes. 

Here, we answer to the following question: what are the implications of social media platforms competing for users' attention on opinion dynamics? To address this, we develop a model of opinion dynamics which integrates three main ingredients. i) People can connect on different platforms, each platform represented by a layer in a \emph{multiplex} network. These networks differ in their recommendation algorithms (a single parameter representing its homophily) and in their political focus. ii) Users' opinions evolve according to a well-known non-linear model \cite{baumann2020modeling}, on the basis of interactions taking place on the multiplex. Opinions depend on the \emph{social interaction strength}, issue controversy and the heterogeneous activity profile on social media. iii) Users allocate their time among the different social media platforms, depending on their personal preferences and limited information processing.

Our model provides two pivotal insights. Firstly, when considering two platforms --- one governed by a neutral recommendation algorithm and the other by a homophily-centric algorithm --- we find that even with a user majority on the neutral platform, opinion polarization can endure. Secondly, as users dynamically allocate their social engagement across platforms based on their (strong or weak) homophilic inclinations, agents manifest a pronounced separation across platforms. While most associate network fragmentation with ``echo chambers'',  this emergent multi-platform segregation boosts user satisfaction without increasing polarization. These findings highlight the importance of recognizing the multifaceted avenues through which individuals assimilate information.

\section{Results}
\begin{figure}[ht]
    \centering
    \includegraphics[scale=0.8]{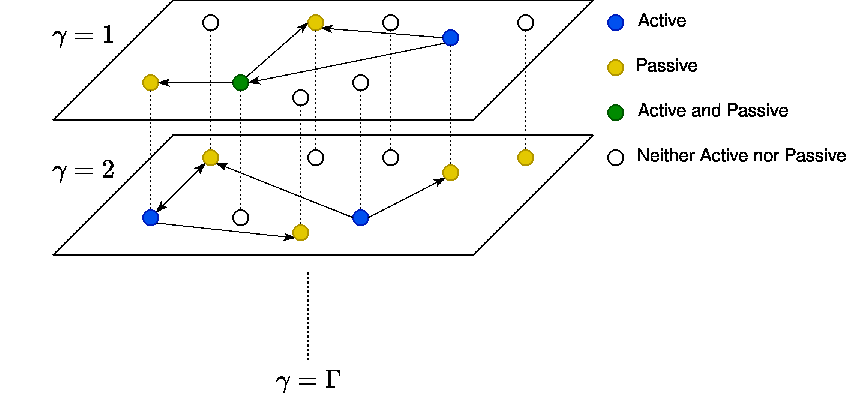}
    \caption{Opinion dynamics on a multiplex network. Dashed lines indicate that the same nodes (users) are shared among the layers. In this example, every active user (blue dots) contacts $m=2$ passive users (yellow dots). A passive user can reciprocate a link with probability $r$ (see Sec.~\ref{sec:model}). A user can be both active and passive on the same platform (green dots), or not engaging at all in social media activity (white dots). }
    \label{fig:multiplex}
\end{figure}

\sloppy
As detailed in Sec.~\ref{sec:model}, we consider a system composed by $\Gamma$ social platforms populated by $N$ agents (see Fig.~\ref{fig:multiplex}) having continuous opinions $x_i(t)\in (-\infty,+\infty), i=\{1,...N\}$. The opinions evolve according to  $\dot{x_i}=-x_i+\sum_{\gamma=1}^{\Gamma}\big(  K^{(\gamma)}\sum_{j=1}^{N} A_{ij}^{(\gamma)}(t)\tanh{(c x_j)}\big)$, a generalization of a well-known model \cite{Baumann2021}. In absence of social interactions, i.e. the second addend on the r.h.s. equals zero, all opinions relax to $x^*=0$. This assumption allows us to study the network influence on opinions, isolating it from the other possible causes of polarization such as identity politics \cite{stewart2021inequality}, cognitive biases \cite{jonas2001confirmation, wang2020public} and economic inequality \cite{duca2016income}. The $\tanh(cx)$ term reflects the fact that the opinion change for each interaction is limited. The parameter $c$ represents how \emph{controversial} a topic is. For $c$ high enough, all opinions equally contribute to the dynamics since $|\tanh(cx)|$ saturates to $1$, meaning that people are maximally susceptible to be socially influenced. On the other hand, if $c$ is very small, only users with extreme opinions are effectively able to influence others. The platform-dependent parameter $K^{(\gamma)}$ represents the social interaction strength. The larger $K^{(\gamma)}$, the larger opinions change as a consequence of given interactions. Its platform-dependence captures the idea that, for a given topic, users may consider a platform more ``appropriate'' than another. For instance, since the exposure to political content on Facebook is often incidental \cite{nyhan2023like}, while users tend to consume more political news on Twitter \cite{bestvater2022politics, liedke2022social}, it is reasonable to assume that, in the realm of politics, Facebook social interaction strength $K^{(FB)}$ is lower than Twitter $K^{(TW)}$. This results in users giving less credit to the political news they are exposed on Facebook. 

The opinion dynamics model evolves on a multiplex directed network, where each layer represents a single social platform, as pictured in Fig~\ref{fig:multiplex}. At every time step (see Sec.~\ref{sec:model} and SI for further details), user $i$ can be \emph{active} (news producer), \emph{passive} (news consumer) or both with probability $a_i$, $p_i$ and $a_ip_i$ respectively. Conditional on being active (resp. passive), user $i$ chooses a platform $\gamma$ with probability $\rho_i^{(\gamma)}$. Note that he might be active on platform $\gamma$ and passive on platform $\gamma'$, each with probability $\rho_i^{(\gamma)}$ and $\rho_i^{(\gamma')}$, respectively. Active users on a platform/layer contact passive users on the same platform/layer. The probability that a $\gamma$-active user $i$ contacts a $\gamma$-passive user $j$ on platform $\gamma$ at time $t$ is $q_{ij}^{(\gamma)}(t) \propto |x_i(t) - x_j(t)|^{-\beta^{(\gamma)}}$, leading to $A_{ji}^{(\gamma)}(t)=1$. The exponent $\beta^{(\gamma)}$ represents the degree of homophily of the recommendation engine of platform $\gamma$.

\begin{figure}[ht]
    \centering

    \subfigure[]{
        \includegraphics[width=0.3\linewidth]{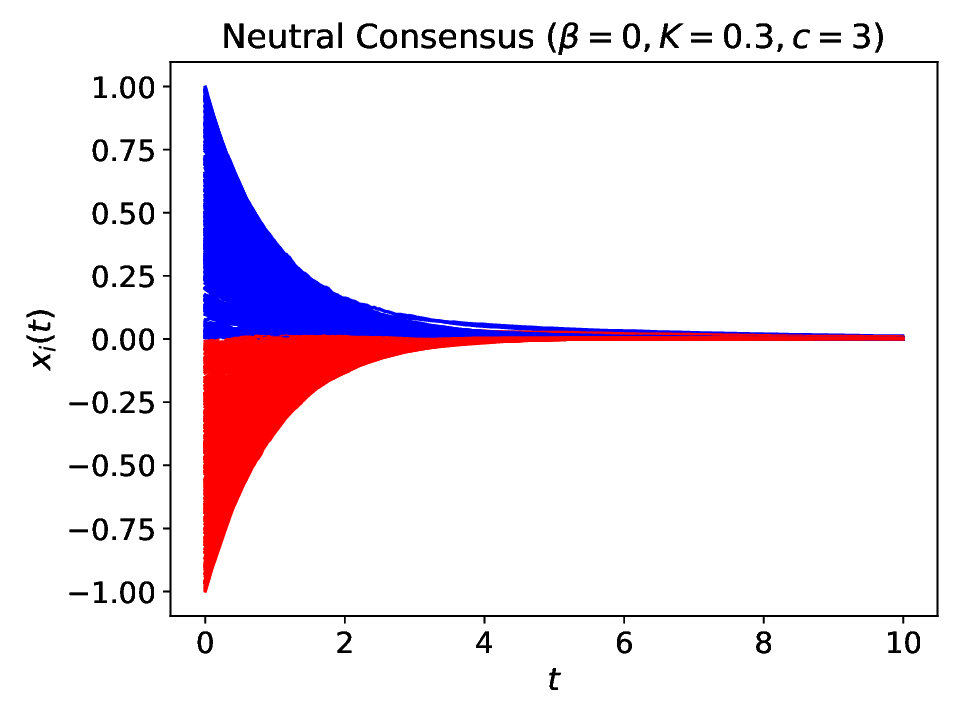}
        \label{fig:subfig1}
    }
    \hfill
    \subfigure[]{
        \includegraphics[width=0.3\linewidth]{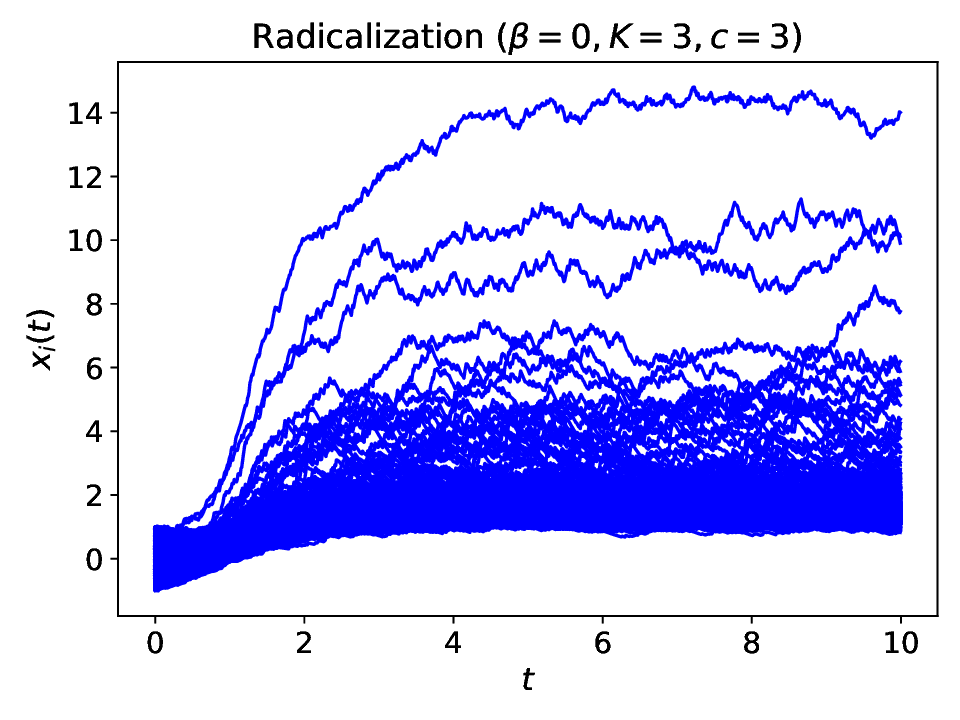}
        \label{fig:subfig2}
    }
    \hfill
    \subfigure[]{
        \includegraphics[width=0.3\linewidth]{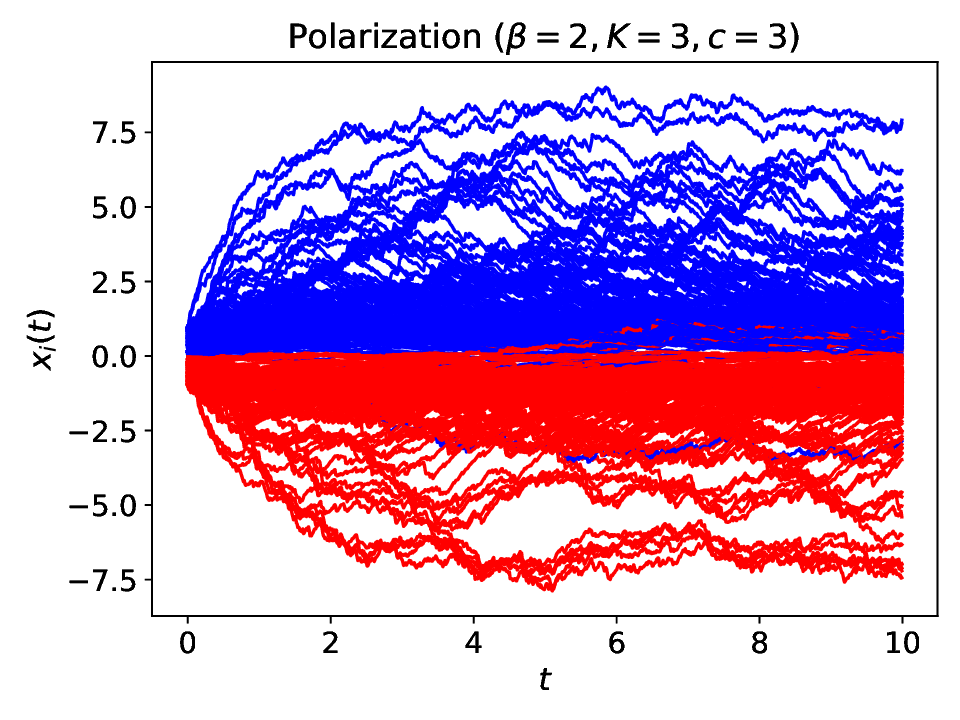}
        \label{fig:subfig3}
    }

    \caption{ (a) Neutral consensus, all opinions converge to zero ($K=0.3, \beta=0$). (b) (One-sided) radicalization ($K=3, \beta=0)$. (c) Opinion polarization, in which opinions split into two opposite sides ($K=3, \beta=2$). Topic controversiality and reciprocity were set to $c=3$ and $r=0.5$.}
    \label{fig:Fabian_main}
\end{figure}

The model with a single platform ($\Gamma=1$) has been studied in \cite{Baumann2021}. Figure~\ref{fig:Fabian_main} shows the main qualitatively different dynamics that the model exhibits, as a function of the different parameters, obtained initializing the opinions uniformly in $[-1,+1]$. Specifically, when $K^{(1)}=K$ is small, social interaction is negligible and opinions relax to $0$ (Fig.~\ref{fig:subfig1}). If instead social coupling is relevant ($K=3$), but no homophilic recommendation engine is present ($\beta=0$), a one-side radicalization appears where all the opinions have the same sign (Fig.~\ref{fig:subfig2}). When both $K$ and $\beta$ are big enough, opinions split into two opposite sides. The intuition is that with $\beta \neq 0$ agents tend to connect only with like-minded peers, an interaction which further polarizes users' stance. This effect makes it even more likely to connect with like-minded individuals; this vicious cycle fosters polarization (a more exhaustive phase diagram for the single-platform model is reported in \cite{Baumann2021}). Note that this phase manifests only if opinions are initialized with different signs. Otherwise, there is no range of parameters which leads to polarization.

\begin{figure}[ht!]
    \centering

    \subfigure[]{
        \includegraphics[width=0.5\linewidth]{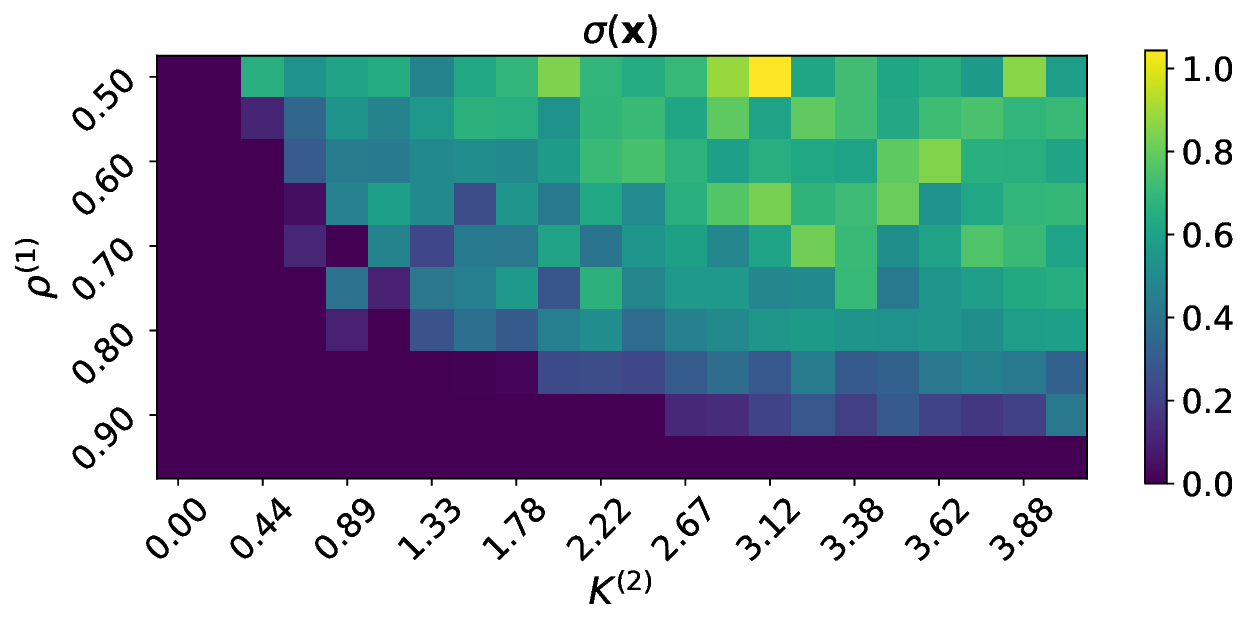}
        \label{fig:std_1}
    }
    \hfill
    \subfigure[]{
        \includegraphics[width=0.5\linewidth]{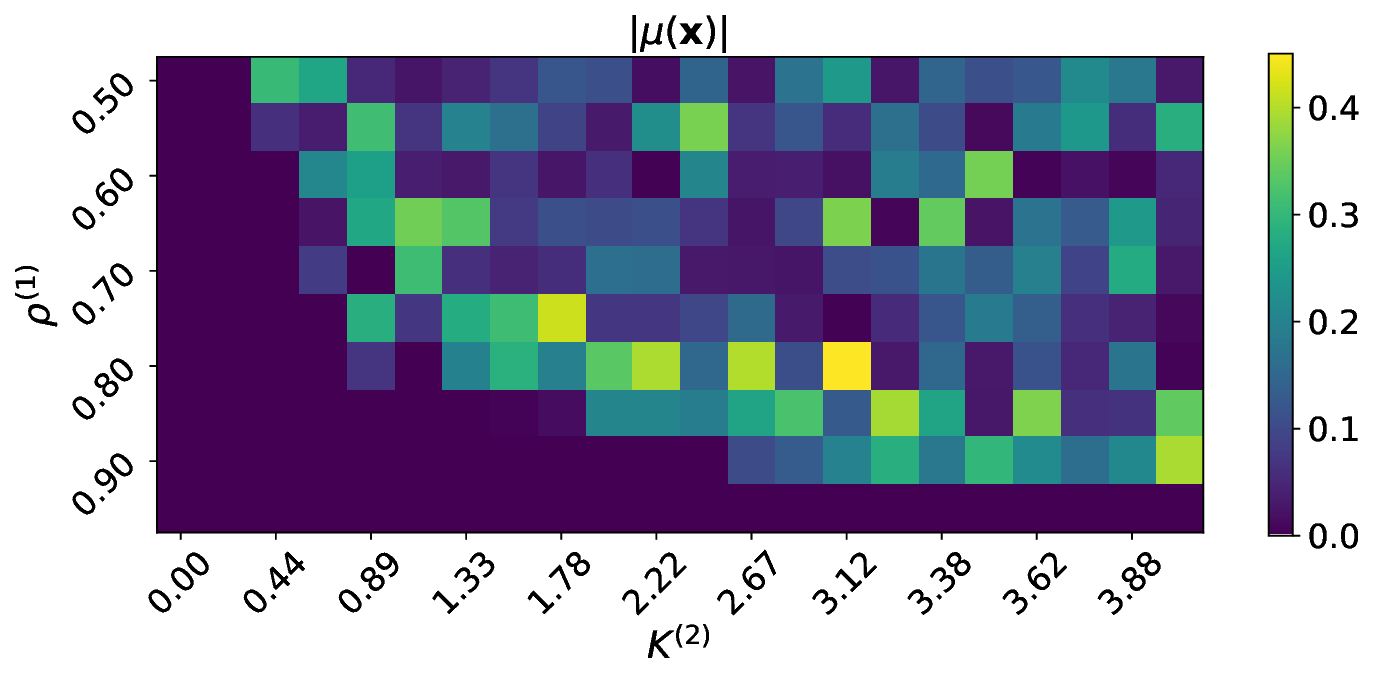}
        \label{fig:avg_1}
    }
    \hfill
    \subfigure[]{
        \includegraphics[width=0.5\linewidth]{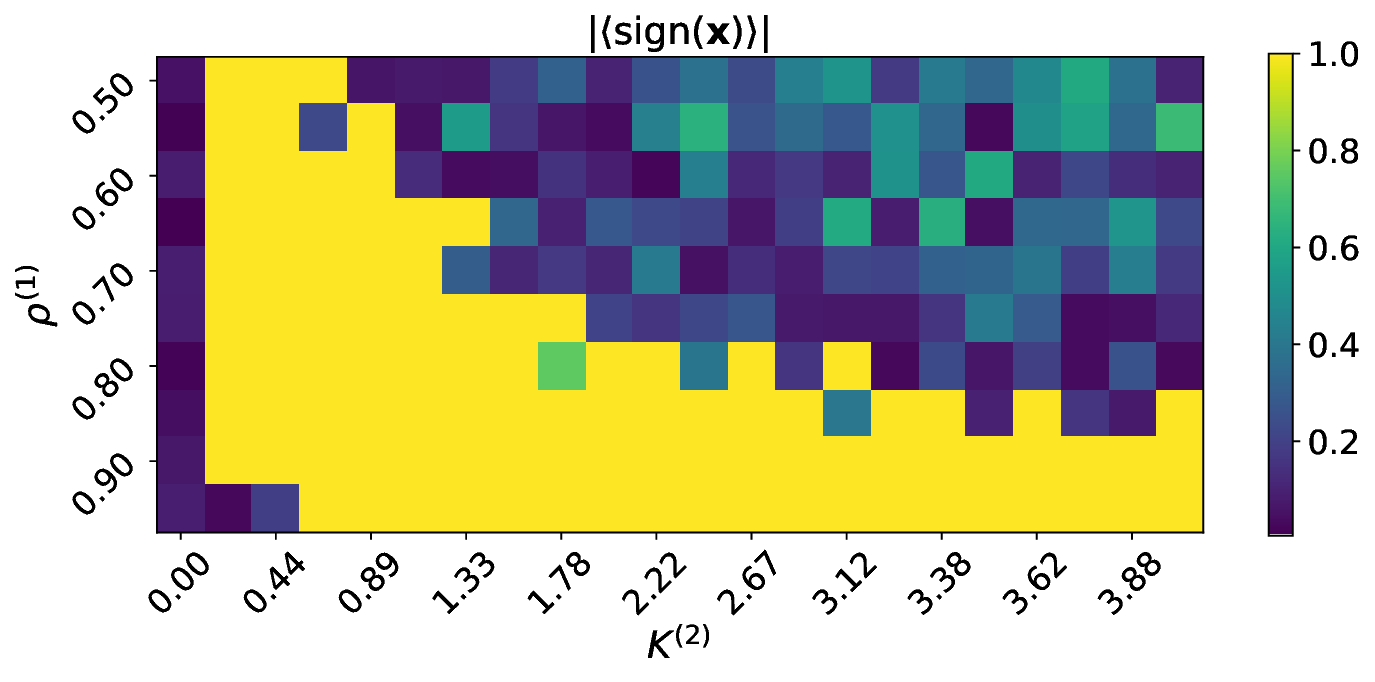}
        \label{fig:sign_1}
    }
    \hfill
    \subfigure[]{
        \includegraphics[width=0.5\linewidth]{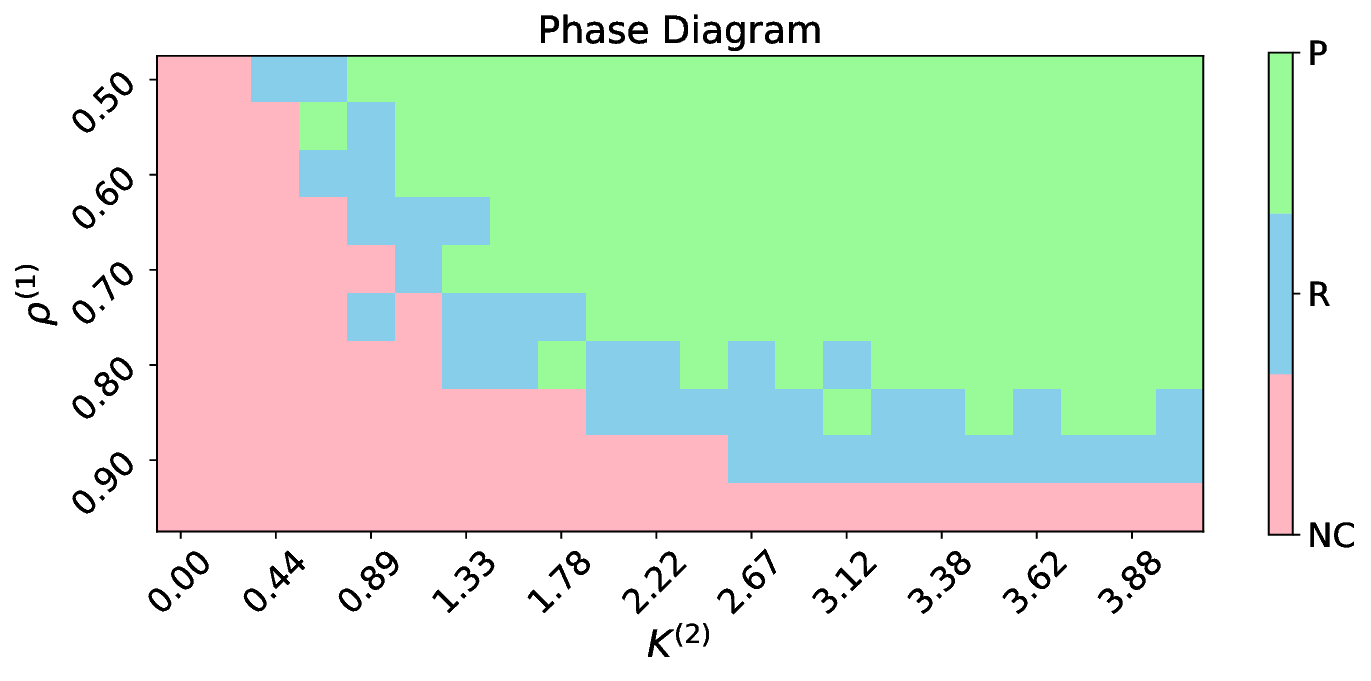}
        \label{fig:phase_dia}
    }
    \caption{Allocation diagram. In all the four panels, on the horizontal axis the social interaction strength of the polarizing platform, and on the vertical axis the time spent on the neutral platform. (a) Standard deviation of the opinions; (b) Absolute value of average opinion; (c) Absolute value of average opinions' sign; (d) Phase diagram. Panel (d) is obtained through ``filtering'' panels (a)-(b)-(c) according to the conditions detailed in the main text. In short: i) NC corresponds to $\sigma(\mathbf{x}) \approx 0$ and $ \mu(\mathbf{x}) \approx 0$). ii) R corresponds to $|\mu(\mathbf{x})|>>0$ and $\langle\mathrm{sign}(\mathbf{x}) \rangle=1$.  iii) P corresponds to $\sigma(\mathbf{x})>>0$ and $\langle\mathrm{sign}(\mathbf{x}) \rangle<1$.}
    \label{fig:rhofisse}
\end{figure}

As anticipated in Sec.~\ref{sec:intro}, we ask whether polarization can persist when users spend a tiny fraction of their time on politically-oriented social media with an homophilic recommendation engine, while engaging the rest of their time on a politically neutral platform. To explore this scenario, we consider $\Gamma=2$ platforms and stationary and homogeneous allocation probabilities, i.e. $\rho_i^{(\gamma)}(t)=\rho^{(\gamma)}$ for all $i$ and for $\gamma=\{1,2\}$.  Clearly, $\rho^{(1)}+\rho^{(2)}=1$. Both the assumptions of stationarity and homogeneity will be later relaxed. Platform $1$ has a set of parameters such that, if users were only there, opinions would converge to neutral consensus ($\beta^{(1)}=0$ and $K^{(1)}$ small). On the other hand, platform $2$ is assumed to adopt an homophilic recommendation engine, which translates in $\beta^{(2)} \neq 0$. The social interaction strength $K^{(2)}$ is left as a varying parameter, meaning that platform $2$ could have exhibited both neutral consensus or polarization if it were the only platform, depending on its value. We are then interested in exploring the opinion dynamics for different values of $\rho^{(1)}$ and $K^{(2)}$. The former represents how long users engage on platform $1$ (the ``politically neutral'' platform); the latter captures how ``polarizing'' platform $2$ is.  We define the \emph{rescaled} vector of opinions at equilibrium as $\mathbf{x}=\{\frac{x_1^{(eq)}}{K^{(1)}\rho^{(1)}+K^{(2)}\rho^{(2)}},\dots,\frac{x_N^{(eq)}}{K^{(1)}\rho^{(1)}+K^{(2)}\rho^{(2)}} \}$ in order to compute a set of three metrics which allow us to distinguish different opinion phases. In particular, such metrics are the standard deviation of the opinions $\sigma(\mathbf{x})$, the absolute value of the average opinion $|\mu(\mathbf{x})|$ and the absolute value of the average opinions' sign $|\langle \mathrm{sign}(\mathbf{x}) \rangle|$. In Figure~\ref{fig:rhofisse} we show the results of our analysis for $\beta^{(1)}=K^{(1)}=0$ and $\beta^{(2)}=3$. We can observe three main phases. i) Neutral Consensus ($\sigma(\mathbf{x}) \approx 0, \mu(\mathbf{x}) \approx 0$), observable when platform $2$ is not polarizing enough (i.e. $K^{(2)}$ not high enough) w.r.t. the time spent on the neutral platform $1$\footnote{The neutral consensus phase can have both $|\langle \mathrm{sign}(\mathbf{x}) \rangle| \approx 0$ and $|\langle \mathrm{sign}(\mathbf{x}) \rangle| \approx 1 $, as the opinions never really reach exactly $0$.}. ii) Radicalization ($|\mu(\mathbf{x})|>>0, \langle\mathrm{sign}(\mathbf{x}) \rangle=1$) is evident by the fact that all opinions share the same sign. Such phase is driven by an initial relaxation towards $x_i(t)\approx 0$ $\forall i$ due to the neutral platform. Then, when close to $0$, opinions start to share the same sign and are progressively amplified by the polarizing (now, rather, radicalizing) platform. iii) Polarization ($\sigma(\mathbf{x})>>0, \langle\mathrm{sign}(\mathbf{x}) \rangle<1$) emerges if platform $2$ can sustain diverging opinions, i.e. if $K^{(2)}$ is big enough to off-set the time users spend on the neutral platform $\rho^{(1)}$. The take home message is that polarization can persist even when users spend most of their time on a politically neutral platform ($\rho^{(1)} >0.5$), thus suggesting the importance of considering that users gather information from different sources. In Fig.~\ref{fig:rhofisse2} of the SI, we show a similar phase diagram for a different value $\beta^{(2)}$. The qualitative picture remains the same.

The above results are obtained by assuming that users' taste for social media platforms may depend on factors which are not captured in the model (e.g. better user interface), therefore we had an homogeneous and stationary allocation probability $\boldsymbol{\rho}=\{\rho^{(1)}, \rho^{(2)}\}$. Hereafter, we suppose that users dynamically \emph{choose} their \emph{Social Media Repertoire} (SMR), depending on the perceived political quality of platforms. Based on the psychological theory of optimal distinctiveness \cite{brewer1991social}, we imagine that each user looks for a trade-off between assimilation (homophily) and differentiation (debate). As detailed in Sec.~\ref{sec:smrupdate}, we capture such desired balance into the (user-dependent) parameter $\phi_i$, which represents the desired fraction of ``far'' opinions (i.e. contributing to differentiation) user $i$ wants to be exposed to. In particular, inspired by bounded confidence theory \cite{hegselmann2002opinion}, user $i$ considers ``far'' opinions those $x$ such that $|x_i-x|>r$. The others are considered ``close'' opinions, contributing to assimilation. Thus, while in bounded confidence theory users do not engage with peers having far opinions, we relax this hypothesis by introducing $\phi_i$, which can be seen as user $i$'s desired probability to contact a distant peer. The idea is to capture the observed desire of debate. Indeed, as reported in \cite{lai2019stance}, despite the general tendency for social media networks to form homogeneous communities, networks formed through reply-to messages reveal a users' stance heterophily, with individuals using replies more often to express divergent opinions. We define the utility (i.e. satisfaction) of each user as $U_i(t)=-(f_i(t)-\phi_i)^2$, where $f_i(t)$ is the fraction of distinctiveness experienced by user $i$, which is compared to the desired one. Clearly, $f_i(t)$ depends on the SMR of user $i$ (i.e. $\boldsymbol{\rho}_i(t)=\{\rho^{(\gamma)}_i(t)\}_{\gamma=1}^{\Gamma}$), since user's exposure to distinctiveness and assimilation depends on the connections formed on each platform, which in turn depend on his allocation probability. We assume therefore that user $i$ dynamically updates $\boldsymbol{\rho}_i(t)$ in order to maximize his $U_i(t)$ (see Sec.~\ref{sec:smrupdate} and SI for additional details).

In the context of platforms' battle for users' attention, social media are interested in maximizing users' satisfaction (which translates in higher activity, thus revenue).
We will now show how a market populated by two platforms achieves higher average satisfaction without the undesired drawback of increasing polarization. This is surprising, as one would expect that the additional degree of freedom increases social fragmentation, thus increasing polarization.

Suppose that each platform can vary its characteristics by tuning its $\beta^{(\gamma)}$\footnote{We assume that $K^{(\gamma)}=K$ for all $\gamma$, implying that they all have the same political focus.}. First, we consider a single platform (i.e. $\Gamma=1$) and consider $K=c=3$, $r=2$ and $\phi_i=0.2$ $\forall i$. In this setting, the highest single-platform average utility $\langle U \rangle_1$ is reached in $\beta^*=3$, i.e. $\langle U \rangle_{\Gamma=1}(\beta^*)=\langle U \rangle_1$. 
\begin{figure}[ht!]
    \centering

    \subfigure[]{
        \includegraphics[width=0.5\linewidth]{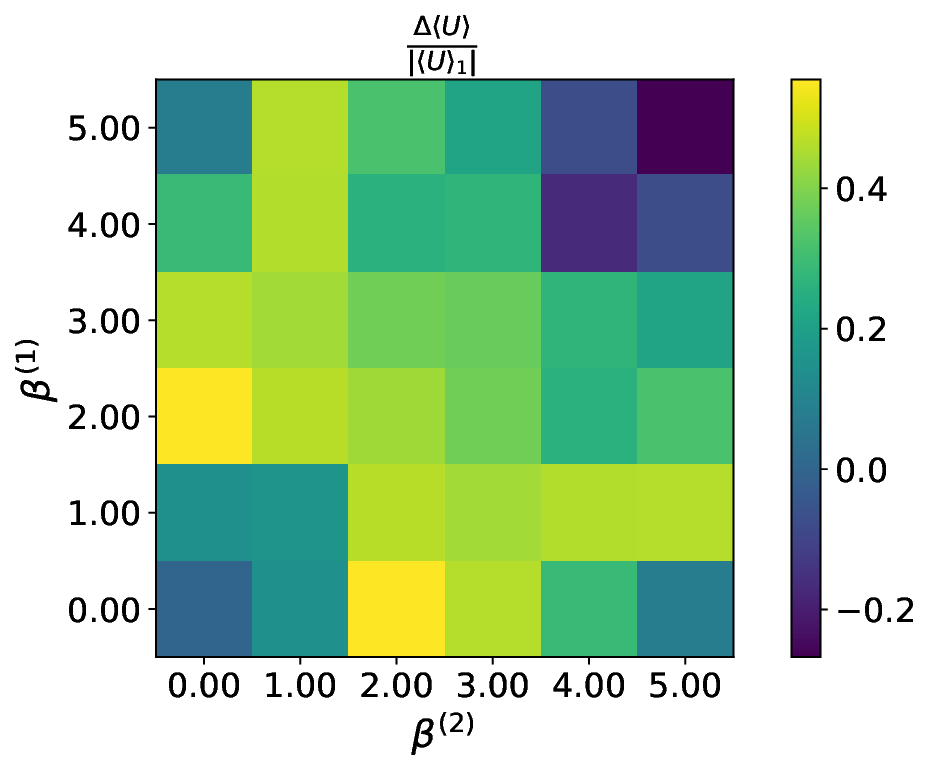}
        \label{fig:u2plat}
    }
    \hfill
    \subfigure[]{
        \includegraphics[width=0.5\linewidth]{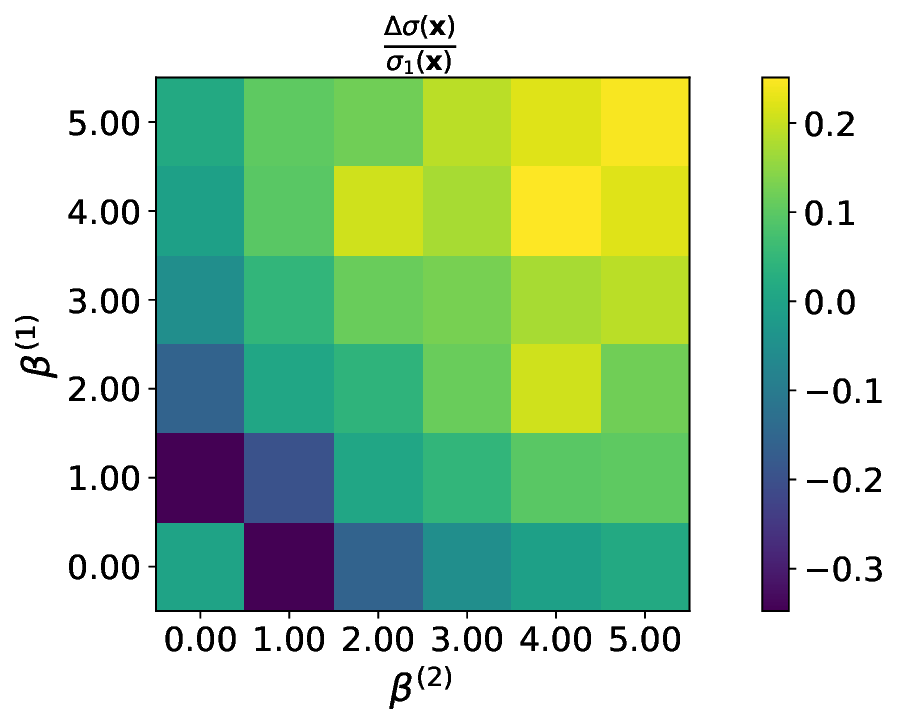}
        \label{fig:std2plat}
    }
    
    \caption{Relative increment passing from $\Gamma=1$ to $\Gamma=2$ platforms of (a) satisfaction and (b) opinions' standard deviation. Neglecting the first row and the last column ($\beta^{(1)}=0$ or $\beta^{(2)}=0$), which correspond to a radicalized regime, for ``moderate'' values of homophilic algorithm (e.g. $\beta^{(1)}=1$ and $\beta^{(2)}=2$), an increased satisfaction ($\approx 40\%$) is clear. Moreover, in such points, the standard deviation relative increse in negligible ($\approx 0\%$). For very high values of homophilic recommendations (e.g. $\beta^{(1)}=5$ and $\beta^{(2)}=5$), the satisfaction drops significantly while the standard deviation increases, making this the worst possible scenario.}
    \label{fig:rhodin2plat}
\end{figure}
On the other hand, a market populated by $\Gamma=2$ platforms can result in an higher satisfaction. Figure~\ref{fig:rhodin2plat} summarizes our results by showing the variation of average utility and standard deviation in passing from $\Gamma=1$ to $\Gamma=2$ for different values of $\beta^{(1)}$ and $\beta^{(2)}$. In particular, we define $\Delta \langle U \rangle(\beta^{(1)},\beta^{(2)}) = \langle U\rangle_{\Gamma=2}(\beta^{(1)},\beta^{(2)})-\langle U \rangle_1$ and $\Delta \sigma (\mathbf{x}; \beta^{(1)},\beta^{(2)}) = \sigma_{\Gamma=2}(\mathbf{x}; \beta^{(1)},\beta^{(2)})-\sigma_1(\mathbf{x})$, where $\sigma_1(\mathbf{x}) = \sigma_{\Gamma=1}(\mathbf{x}; \beta^*)$. Even if in the figure we reported, for the sake of completeness, the values corresponding to $\beta^{(1)}=0$ or $\beta^{(2)}=0$, in the following discussion we are going to neglect those points. The reason is that they correspond to a radicalized regime, which is highly undesired in a democracy, which lies on dialogue and debate. Thus, restricting the attention to$(\beta^{(1)}, \beta^{(2)}) \in [1,5]^2$, the average utility increases for many couples of values with respect to the \emph{best} possible single-platform average utility. The reason is that users can dynamically change their SMR, thus allocating their time among platforms to satisfy their desired differentiation $\phi$. Moreover, focusing our attention on the point $(\beta^{(1)}=2$, $\beta^{(2)}=1)$ (or, by symmetry, $(\beta^{(1)}=1$, $\beta^{(2)}=2)$), which is arguably very close to the two-platform optimum, we can see that the corresponding variation of polarization is almost zero, i.e. $\Delta \sigma(\mathbf{x}; 2, 1) \approx 0$. As we show in Fig.~\ref{fig:seg_grandi_piccoli}, the explanation for higher satisfaction without increasing polarization is the emerging quasi-segregation between moderate and extreme opinions. The moderate individuals, in absence of the extreme ones who populate another platform, get radicalized less, effectively reducing polarization. This is in qualitative agreement with findings on media habits and opinion stance, where individuals at both the left and right ends of the spectrum tend to be clustered around a single media source\footnote{Note however that there are some important differences between liberals and conservatives that the model cannot capture} \cite{prior2013media}. So, by focusing only on users' satisfaction, the competition of platforms brings notable benefits. Finally, the points for which satisfaction decreases with respect to $\Gamma=1$ are those where both $\beta^{(1)}$ and $\beta^{(2)}$ are large enough. In this regime, users connect only to very close peers, and can not find an alternative, cross-cutting platform. Thus, they cannot satisfy their desire for differentiation encapsulated in $\phi=0.2$. In SI Fig.~\ref{fig:rhodin2plat2} we show an alternative situation, where $\phi=0.1$. In that case, users' desire for differentiation is too low and maximizing the utility leads to an increased polarization.

\section{Discussion}

Online social networks have become crucial for political news consumption \cite{bond201261,shearer2019americans}. Existing studies \cite{bakshy2015exposure,del2016spreading,garimella2018political,huszar2022algorithmic,levy2021social,nyhan2023like,holme2006nonequilibrium,wang2020public,santos2021link,baumann2020modeling} explore the impact of recommendation algorithms on opinion dynamics, with some findings on Facebook's limited influence \cite{bakshy2015exposure,nyhan2023like}. However, news consumption spans various platforms \cite{hasebrink2006media}, and individual preferences \cite{zillmann2000mood} play a role. It is thus crucial to consider the interplay between news repertoires and political communication processes to understand political opinions fully.

Here we show that when examining two platforms --- one with a low political focus and a neutral recommendation algorithm, and another more politically oriented with a homophily-based algorithm --- even if users spend the majority of their time on the neutral platform, opinion polarization can persist. This result casts some doubts on the generality of conclusions drawn from the recent study on the impact of Facebook recommendation algorithm on political opinions \cite{nyhan2023like}, which shows that people's political attitudes did not significantly change when they were exposed to more diverse content. Indeed, the consumption of political news on Facebook is incidental (low political focus), and the polarization might originate from (little) time spent on other news sources. 

When we allow users to dynamically optimize their satisfaction by adjusting their SMR, the further segregation of individuals brought by multiple platforms leads to an increase of global satisfaction without, unexpectedly, an increase of polarization, with respect to the single platform case. In fact, more active (i.e. more extreme) users separate from users with ``low'' activity (i.e. more moderate), thus the latter are no longer influenced by the former, resulting in a reduced (or, at least, not increased) polarization.

While our multiplatform opinion dynamics model offers a comprehensive framework for understanding interactions across diverse platforms, it is imperative to acknowledge the inherent constraints and nuances that underpin its design. The following considerations delve into these aspects, shedding light on both the model's robustness and areas that warrant further refinement. In the first place, while our model describes social media platforms, there exist traditional media not captured by it, which influence users' opinions. Broadcasting platforms such as TVs and radios are an example, contributing to opinion formation in a way which is structurally and temporally different from social media. With reference to our model, they would act as hub nodes in the network of interactions, each diffusing a certain opinion to other nodes. However such nodes, unlike ``normal'' ones, do not change (or, rather, change very slowly) their opinions over time. Thus, effectively, traditional platforms can be seen as idiosyncratic relaxation terms in Eqn.~\ref{eq:full_equation}, as opposed to having all opinions relax to zero $x_i(\infty)=x^*=0$ when $K^{(\gamma)}=0$ for all $\gamma$. The interplay between these two sources of information is left for future studies. 

Regarding how users are affected by peers, while we assume that they converge (i.e. their opinions approach each other when interacting), other models \cite{santos2021link,bail2018exposure, shmargad2019partisan, paluck2010better} consider also the presence of ``polarizing'' nodes, whose opinions move away if exposed to those of opposite sign. Since this characterization does not qualitatively change our results, we neglect it. In fact, as we have shown, both polarization and consensus emerge without the introduction of this additional feature. 

Our last assumption is the conservation of activity over time, i.e. $a_i(t)=a_i$ $\forall i,t$. In modeling users' satisfaction across platforms, it would be reasonable to require that users' activity (which corresponds to users' engagement on social platforms) could decrease if they are not able to reach a certain degree of satisfaction (i.e. a certain value of their utility). However, modeling such behavior would require a clear definition of what a decreasing activity really mean --- do users move their attention to traditional media? Do they allocate their SMR on a not-considered social platform? Or do they literally stop consuming news? --- We believe that it is reasonable to assume the considered system (which, we note, can be made arbitrary large by increasing $N$ and/or $\Gamma$) as an \emph{isolated} social system, in which social energy (i.e. users' activity) is conserved\footnote{Although note that daily time spent on social media by internet users worldwide has been steadily increasing from 2012 to 2023\cite{social2023january}.}. In fact, it is possible to think that our model is valid in a span of time $T$ in which activities remain constant, meaning that they change on a time scale much greater than $T$.

Finally, we want to stress how $\beta^{(\gamma)}$ encapsulates the role of the recommendation engine of platform $\gamma$. Typically, recommendation systems aim to infer users' needs, tastes and preferences, on the basis of their behavior on the platform, in order to suggest them the ``best'' product \cite{burke2002hybrid}. In terms of social media platforms, recommendation systems try to connect people who are ``similar'' according to some metric. Obviously, different platforms collect different kind of users' data, deploy different recommendation algorithms and have different purposes. Such dissimilarities translate into a non-homogeneous degree of recommendations' homophily. We imagine that the distances between users, evaluated by the recommendation algorithms by considering the huge amount of data social media collect, can be projected into $1$-dimensional distances between political opinions, further distorted by the degree of homophily $\beta^{(\gamma)}$. Such assumption is based on the well-known phenomenon of issue alignment \cite{baldassarri2008partisans, dellaposta2015liberals, Baumann2021}, i.e. individuals are much more likely to have a certain combination of opinions than others. On this note, an interesting extension of the model would be to consider a multi-topic opinion dynamics on a multiplex network; the idea would be that on a given layer/platform it might be more likely to talk about a topic than another.

\section{Model}
\label{sec:model}
We consider a system of $N$ agents, where each agent $i$ has a one-dimensional continuous opinion variable $x_i(t) \in (-\infty, +\infty)$. The sign of $x_i$ describes the agent’s stance (e.g. being pro or against abortion). The absolute value of $x_i$ quantifies the strength of this opinion: the larger $|x_i|$, the more extreme the stance of agent $i$. Moreover, we also consider that agents can interact on $\Gamma$ different platforms. Each agent allocates his ``time'' in these $\Gamma$ platforms. In the following sections we detail how opinions and interaction networks evolve and how each agent divides his attention among the platforms (see SI for a detailed outline of the model simulation).
\subsection{Opinions update}
The opinion dynamics is driven by the interactions among agents, captured by a system of $N$ coupled ordinary differential equations,
\begin{equation}
\label{eq:full_equation}
    \dot{x_i}=-x_i+\sum_{\gamma=1}^{\Gamma}\Big(  K^{(\gamma)}\sum_{j=1}^{N} A_{ij}^{(\gamma)}(t)\tanh{(c x_j)}\Big). \quad i \in \{1,..N\}
\end{equation}
$K^{(\gamma)} > 0 $ represents the social interaction strength among agents on platform $\gamma$. The $\tanh{(c x)}$, with $ c > 0$, embodies the fact that an agent $i$ influences others in the direction of his own opinion, but such influence is ``bounded''. The term $A_{ij}^{(\gamma)}(t)$ is the entry of the $N\times N$ temporal Adjacency matrix $A^{(\gamma)}(t)$ corresponding to platform $\gamma$. 
\subsection{Network update}
At every time step, user $i$ can actively engage with the social media on platform $\gamma$ with probability $a_i \rho_i^{(\gamma)}$, and/or passively engage on platform $\gamma'$ with probability $p_i \rho_i^{(\gamma')}$ \footnote{Note that we can have $\gamma=\gamma'$.}. Active users contact users that are passive on the same platform, meaning that the opinions of the former affect the opinions of the latter. For example, if $j$ is active on $\gamma$ and contacts $i$, who is passive on $\gamma$, then $A_{ij}^{(\gamma)}=1$. The term $a_i$ (resp. $p_i$) is called \emph{activity} (resp. \emph{passivity}). Moreover, $\rho_i^{(\gamma)}$ represents the probability that user $i$, conditional on being active/passive, chooses platform $\gamma$. Of course, $\sum_{\gamma=1}^{\Gamma}\rho_i^{(\gamma)}=1$. We assume that $a_i \in [\epsilon, 1]$ for all $i$, and that the activities are distributed according to a power law $F(a) \sim a^{-\eta}$. Moreover, we assume for simplicity that $p_i=1$ $\forall i$.

The temporal adjacency matrices $A_{ij}^{(\gamma)}(t)$ are assumed to evolve according to an activity-driven (AD) temporal network \cite{perra2012activity}. At each time step, each $\gamma$-active user contacts $m$ $\gamma$-passive users. It is further assumed that these links are reciprocated with probability $r$. The probability $q_{ij}^{(\gamma)}$ that agent $i$ contacts agent $j$ on platform $\gamma$ is given by the following expression:
\begin{equation} 
\label{eq:probarew}
    q_{ij}^{(\gamma)}=\frac{|x_i-x_j|^{-\beta^{(\gamma)}}}{\sum_{k\in \mathcal{P}^{(\gamma)}}|x_i-x_k|^{-\beta^{(\gamma)}}}
\end{equation}
where $\mathcal{P}^{(\gamma)}$ is the set of $\gamma$-passive users and $\beta^{(\gamma)} \geq 0$ captures the degree of homophily of $\gamma$'s recommendation algorithm.

\subsection{SMR update}
\label{sec:smrupdate}
While for a part of our results we considered $\rho_i^{(\gamma)}=\rho^{(\gamma)}$ homogeneous and constant in time, we also developed a model for which it changes on the basis of the observations of each user. In particular, we suppose that users allocate their ``social energy'' among platforms depending on their perceived quality. Grounded on the well-known psychological theory of optimal distinctiveness \cite{brewer1991social}, individuals desire a balance of between assimilation (homophily) and differentiation (debate)(see \cite{chu2021heterogeneity} for an example of discrete opinion dynamics with optimal distinctiveness preferences). Borrowing from bounded confidence theory \cite{krause2000discrete}, an agent with opinion $x$ considers those with opinion in $[x-r, x+r]$ contributing to assimilation, while the others to differentiation. Formally, if $\alpha_i^{(\gamma)}(t)$  (resp. $\delta_i^{(\gamma)}(t)$) is the number of in-degree connections contributing to assimilation (resp. differentiation) on $\gamma$ at time $t$ for user $i$\footnote{$\alpha_i^{\gamma}(t)+\delta_i^{\gamma}(t)$ is always equal to the total in-degree of node $i$. }, we define his utility as: 
\begin{align}
    \begin{split}
    U_i(t_n)&=-(f_i(t_n)-\phi_i)^2 \\
    &=-\Bigg(\frac{\sum_{\gamma}\sum_{m=n-L+1}^n\delta_i^{(\gamma)}(t_m)}{\sum_{\gamma}\sum_{m=n-L+1}^n\delta_i^{(\gamma)}(t_m)+\sum_{\gamma}\sum_{m=n-L+1}^n\alpha_i^{(\gamma)}(t_m)}-\phi_i\Bigg)^2.
    \end{split}
\end{align}
Here $f_i(t)$ is the experienced distinctiveness, while $\phi_i$ the desired one. The parameter $L$ encapsulates the time window over which users perceive the distinctiveness (which, in time units, is $\tau=Ldt$), thus representing their ``memory''. For example, consider $\Gamma=2$, $L=1$ and $\phi=0.5$. If user $i$ is connected to user $j$ on platform $1$ with $|x_j-x_i|<r$, and to user $k$ on platform $2$ with $|x_k-x_i|>r$ at time $t_n$, then it follows that $\delta_i^{(1)}(t_n)=\alpha_i^{(2)}(t_n)=0$ and $\delta_i^{(2)}(t_n)=\alpha_i^{(1)}(t_n)=1$. Thus, $f_i(t_n)=\phi_i=0.5$ and $U_i(t_n)=0$ is maximized. 

It remains to specify how users maximize their utility. They can only control the fraction of time spent on each platform $\gamma$, proportional to $\rho_i^{(\gamma)}(t_n)$. We assume that users update their platform allocation every $L$ steps, i.e. their preferences stay constant during the time interval over which assimilation and differentiation are experienced\footnote{In other words, users gather experience before changing their mind about a given social media.}. For this reason, each user can estimate the quality of platforms \emph{given} her current preferences. Such estimates are in turn used to update the platform allocation, on the basis of an anticipated utility. It is reasonable to assume that each user acts as if the assimilation and differentiation experienced on each platform are proportional to the time spent on it. For this reason, it is possible to write:
\begin{align}
\begin{split}
\label{eq:slopes}
    &\sum_{m=n-L+1}^n \delta^{(\gamma)}_i(t_m) =L \omega_{\delta_i}^{(\gamma)}(t_n)\rho_i^{(\gamma)}(t_n),\\
    &\sum_{m=n-L+1}^n \alpha^{(\gamma)}_i(t_m) = L\omega_{\alpha_i}^{(\gamma)}(t_n)\rho_i^{(\gamma)}(t_n).
    \end{split}
\end{align}
Here, $\omega_{\delta_i}^{(\gamma)}(t_n)$
and  $\omega_{\alpha_i}^{(\gamma)}(t_n)$ are the differentiation and the assimilation slopes estimated by the user $i$.
Formally, defining $\boldsymbol{\rho}=(\rho^{(1)},\dots,\rho^{(\Gamma)})$, each user aims to maximize:
\begin{equation}
\label{eq:estimated_utility}
    \hat{U}_i(\boldsymbol{\rho}, t_n)=-\Bigg(\frac{\boldsymbol{\omega}_{\delta_i}(t_n)\cdot \boldsymbol{\rho}}{\boldsymbol{\omega}_{\delta_i}(t_n)\cdot \boldsymbol{\rho}+\boldsymbol{\omega}_{\alpha_i}(t_n)\cdot \boldsymbol{\rho}}-\phi_i\Bigg)^2,
\end{equation}
where $\boldsymbol{\omega}_{\delta_i}(t_n) = (\omega^{(1)}_{\delta_i}(t_n),\dots,\omega^{(\Gamma)}_{\delta_i}(t_n))$  is the vector of differentiation slopes estimated using user's previous interactions (an analogous definition holds for the assimilation slopes $\boldsymbol{\omega}_{\alpha_i}(t_n)$). $\hat{U}_i(\boldsymbol{\rho}, t_n)$ is the utility \emph{estimated} by user $i$ at time $t_n$. He assumes that it represents his future satisfaction (i.e. for $t>t_n$) depending on how he reallocates his $\boldsymbol{\rho}_i(t_{n+1})$. His assumption lies on hypothesizing that the slopes $\boldsymbol{\omega}_{\alpha_i}$ and $\boldsymbol{\omega}_{\delta_i}$ are roughly constant in time (which, instead, can vary due to the reallocation of all the other agents). User $i$ updates then according to:
\begin{equation}
\boldsymbol{\rho}_i(t_{n+1})=
\begin{cases}
    \boldsymbol{\rho}_i(t_n) &\enspace \mathrm{if} \enspace n/L \notin \mathbb{N}\\ 
\label{eq:argmax}
    \argmax_{\boldsymbol{\rho}} \hat{U}_i(\boldsymbol{\rho}, t_n) &\enspace \mathrm{otherwise}.
    \end{cases}
\end{equation}
Clearly, the utility of user $i$ depends not only on his $\boldsymbol{\rho}_i$, but also on how the other users have allocated their time on social media.  Let us stress that in our model, users can only decide whether to be on a particular platform, but the connections are decided entirely by the recommendation algorithm. A justification for this comes from the work of \cite{viswanath2009evolution}, which shows that on Facebook the number of new links per day increased abruptly after the introduction of a ``who to follow'' recommendation algorithm. In other words, the individual agency in choosing connections is negligible with respect to the volume of content suggested by the platform itself. 

\subsection{Combined dynamics}
\label{sec:comb_dyn}
We focus on a regime in which the three processes described above have different time scales. As already mentioned, we consider the network dynamics being much faster than the opinion dynamics. This is especially true in online social media context. In particular, for each network update we integrate Eq~\eqref{eq:full_equation} for $dt=0.001$. Moreover, the SMR dynamics lies between the two, representing the fact that the choice of the allocation among platforms is faster than the opinion dynamics, but requires a significant number of observations and interactions. Specifically, each user updates his preferences $\boldsymbol{\rho}_i(t_n)$ every $L=100$ time steps. In short, every $L=100$ network updates each user modifies her allocation preferences, and every $1000$ network updates the opinions evolve of a unit time.
\section{Conclusions}
Our study examines the impact of social media competition for users' engagement on opinion dynamics. First, we show that opinion polarization can persist as long as users spend a fraction of their time on a homophilic platform, highlighting the importance of multi-sources news diets. Second, we show that individual users' preferences interact in a non-trivial way with the recommendation algorithms in the presence of multiple platforms. The model indeed predicts the observed relationship between news outlet preference and political ideology. Interestingly, a multi-platform setup may be used to curb polarization while keeping user engagement intact. To this end, it is paramount to experimentally investigate users preferences for diversity (estimates for $\phi$), either via surveys or controlled experiments. 
This avenue may help to shed light on healthy synergies between different social media platforms. From the revenue point of view, synergies are already well-known in this environment (think about users spending time on Whatsapp, Instagram and Facebook without ever going out from the Meta universe).
Future research and efforts should thus gather cross-platform data, via survey \cite{diehl2019multi,liedke2022social} or by experiments, in order to fully comprehend the subtle mechanism of opinion formation in online environments. 
\section*{Acknowledgments}
The work of D.G. has been supported by the European Union -- Horizon 2020 Program
under the scheme ``INFRAIA-01-2018-2019 -- Integrating Activities for
Advanced Communities'', Grant Agreement n. 871042, ``SoBigData++: European
Integrated Infrastructure for Social Mining and Big Data Analytics'' \url{http://www.sobigdata.eu}, by the NextGenerationEU -- National Recovery and Resilience
Plan (Piano Nazionale di Ripresa e Resilienza, PNRR) -- Project: ``SoBigData.it -- Strengthening the Italian RI for Social Mining and Big Data Analytics''
-- Prot. IR0000013 -- Avviso n. 3264 del 28/12/2021. A.S. and D.G. acknowledge support from the Dutch Econophysics Foundation (Stichting Econophysics, Leiden, the Netherlands).
G.M.F. acknowledges support from Swiss National Science Foundation (grant number P500PS-211064).

We thank Fernando Santos, Yphtach Lelkes and Keena Lipsitz for enriching comments.

\bibliographystyle{unsrt}
\bibliography{main}

\clearpage

\begin{appendices}
\setcounter{equation}{0}
\setcounter{figure}{0}
\renewcommand{\theequation}{S\arabic{equation}}
\renewcommand{\thefigure}{S\arabic{figure}}
\section{Pseudo code} \label{sec_PsC}
In each time step of the numerical algorithm the opinions, the temporal matrix and the SMR are update according to the following steps:
\begin{enumerate}
    \item Each user $i$ is only active with probability $a_i(1-p_i)$, only passive with probability $(1-a_i)p_i$, active and passive with probability $a_ip_i$ and inert with probability $(1-a_i)(1-p_i)$.
    \item If active (resp. passive), user $i$ chooses the platform $\gamma$ (resp $\gamma'$) on which he actively (resp. passively) engages with probability $\rho^{(\gamma)}_i(t)$ (resp. $\rho^{(\gamma')}_i(t)$) Note that it could be $\gamma=\gamma'$.
    \item If active on platform $\gamma$, agent $i$ influences $m$ distinct agents $j \in \mathcal{P}^{(\gamma)}$ --- where $\mathcal{P}^{(\gamma)}$ is the set of users passively engaged on platform $\gamma$ --- chosen according to Eqn.~\eqref{eq:probarew}. This influence is expressed by updating the temporal adjacency matrix $A_{ji}^{(\gamma)}(t_n) = 1$.
    \item With probability $r$ the directed link  is reciprocated, so that agent $i$ receives influence from $j$, i.e. $A^{(\gamma)}_{ij}(t_n) = 1$.
    \item Opinions $x_i$ are updated by numerically integrating Eq.~\eqref{eq:full_equation} using the total adjacency matrix elements $A_{ij}(t_n)=\sum_{\gamma=1}^{\Gamma} A^{(\gamma)}_{ij}(t_n)$.
    \item Each user $i$ collects the experienced assimilation and differentiation on each platform at the time step $t_n$ as: 
    $$\alpha^{(\gamma)}_i(t_n)=\sum_{j\mathrm{\;s.t.\;}|x_i-x_j|<r}A_{ij}^{(\gamma)}(t_n)$$
    $$\delta^{(\gamma)}_i(t_n)=\sum_{j\mathrm{\;s.t.\;}|x_i-x_j|>r}A_{ij}^{(\gamma)}(t_n).$$
    \item If $\mod(t_n,L)\neq 0$, then the SMR remains constant for all users $\boldsymbol{\rho}_i(t_{n+1})=\boldsymbol{\rho}_i(t_{n}) \enspace \forall i$. If $\mod(t_n,L)=0$, each user updates his SMR according to the following steps:
    \begin{enumerate}
  
        \item estimate differentiation and assimilation slopes $\boldsymbol{\omega}_{\delta_i}(t_n)$ and $\boldsymbol{\omega}_{\alpha_i}(t_n)$ according to Eqn.~\eqref{eq:slopes}
        \item updates SMR according to $\boldsymbol{\rho}_i(t_{n+1})=\argmax_{\boldsymbol{\rho}}\hat{U}_i(\boldsymbol{\rho},t_n)$, where $\hat{U}_i(\boldsymbol{\rho},t_n)$ is defined in Eq.~\eqref{eq:estimated_utility}. To perform the utility maximization, we use a gradient descent algorithm on the $\Gamma$-dimensional simplex, consisting of $n_{GD}$ iterations of learning rate $\Delta_{GD}$. In particular, the following equation is iterated $n_{GD}$ times: $$\boldsymbol{\rho}_i(k+1)=\mathbf{P}_{\Gamma}\big(\boldsymbol{\rho}_i(k)-\Delta_{GD}\boldsymbol{\nabla}\hat{U}_i(\boldsymbol{\rho}_i(k), t_n) \big),$$ where $\mathbf{P}_{\Gamma}$ is the projection on the $\Gamma$-dimensional simplex, $k$ runs from $k=0$ to $k=n_{GD}-1$, $\boldsymbol{\rho}_i(0)=\boldsymbol{\rho}_i(t_n)$ and $\boldsymbol{\rho}_i(n_{GD})=\boldsymbol{\rho}_i(t_{n+1})$.
  \end{enumerate}
    \item After each time step the temporal networks $A_{ij}^{(\gamma)}(t_n)$ are deleted.
\end{enumerate}
Of course, when we considered a homogeneous and stationary (HS) SMR, steps $6$. and $7$. are ignored, i.e. $\boldsymbol{\rho}_i(t_n)=\boldsymbol{\rho}_i(t_0)$ $\forall i,t$. As done in \cite{baumann2020modeling}, we integrate Eq.~\eqref{eq:full_equation} using an explicit fourth-order Runge-Kutta method with a time step of $dt = 0.01$ in the case of HS SMR, and $dt=0.001$ in the case of evolving SMR. In the latter, we also consider $L=100$. This leads to a timescale separation between the network dynamics, the SMR update and the opinion evolution mentioned in Sec.~\ref{sec:comb_dyn}.

We independently sampled activities $\{a_i\}_{i=1}^N$ from the power law $F(a)=(1-\eta)a^{-\eta}/(1-\epsilon^{1-\eta})$, with parameters $\eta=2.1$ and $\epsilon=0.01$ \cite{baumann2020modeling}. We also set $p_i=1$ $\forall i$. Moreover, we consider the following parameters' values: $N=800$, $r=0.5$ and $c=3$.

Throughout all of our simulations, we start with initial opinions $\{x_i(0)\}_{i=1}^N$ uniformly distributed in $[-1,1]$. In the case of dynamical SMR, we let the opinions evolve for $n_{boot}=2000$ steps before allowing users to reallocate themselves (i.e. we ignore point $7$. for the $n_{boot}$ steps before $t_0$). The reason is that we want to provide as input to our SMR update model an opinions' spectrum which is at (metastable, in the case of polarization) equilibrium . We initialize $\boldsymbol{\rho}_i(t_n)=1/\Gamma$ $\forall n \in \{-n_{boot}+1,\dots, 0\}$.  Then, at $t_0$, we ``turn on'' the SMR allocation described in point $7$. of the pseudo-code.

\section{Robustness of the phase diagram}
\begin{figure}[ht!]
  \centering
  \includegraphics[width=0.8\linewidth]{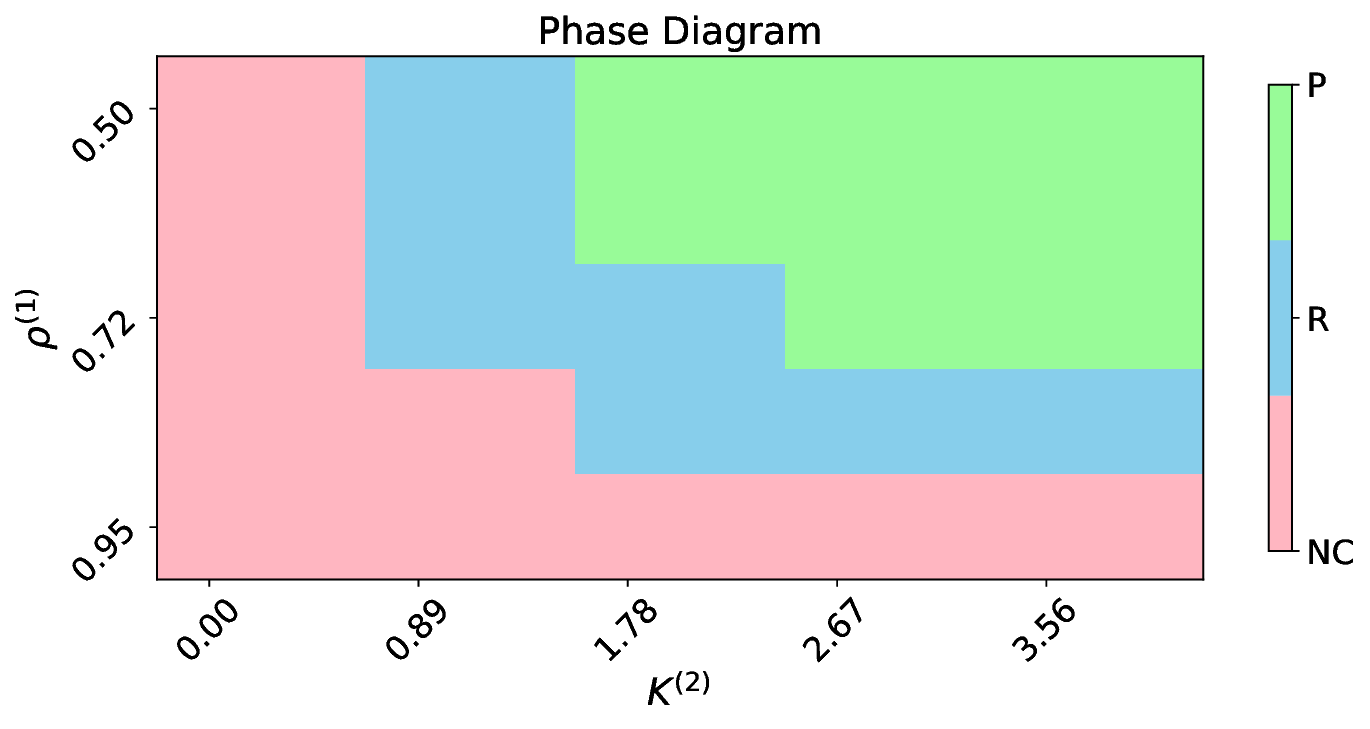}
  \caption{Phase diagram in the case $\beta^{(1)}=0, \beta^{(2)}=2$. The other parameters are as in the corresponding Fig.~\ref{fig:rhofisse} reported in the main text.}
  \label{fig:rhofisse2}
\end{figure}

Here we show the phase-diagram reported in the main text with $\beta^{(2)}=2$, i.e. assuming the polarizing platform provides more cross-cutting content. By comparing Fig. \ref{fig:rhofisse} and Fig. \ref{fig:rhofisse2}, we can see that the green region shrinks as $\beta^{(2)}$ decreases, in accordance with the meta-stability analysis of polarization reported in \cite{baumann2020modeling}.
\section{Multi-platform segregation} \label{sec_seg}
\begin{figure}[ht]
  \centering
  \includegraphics[width=0.8\linewidth]{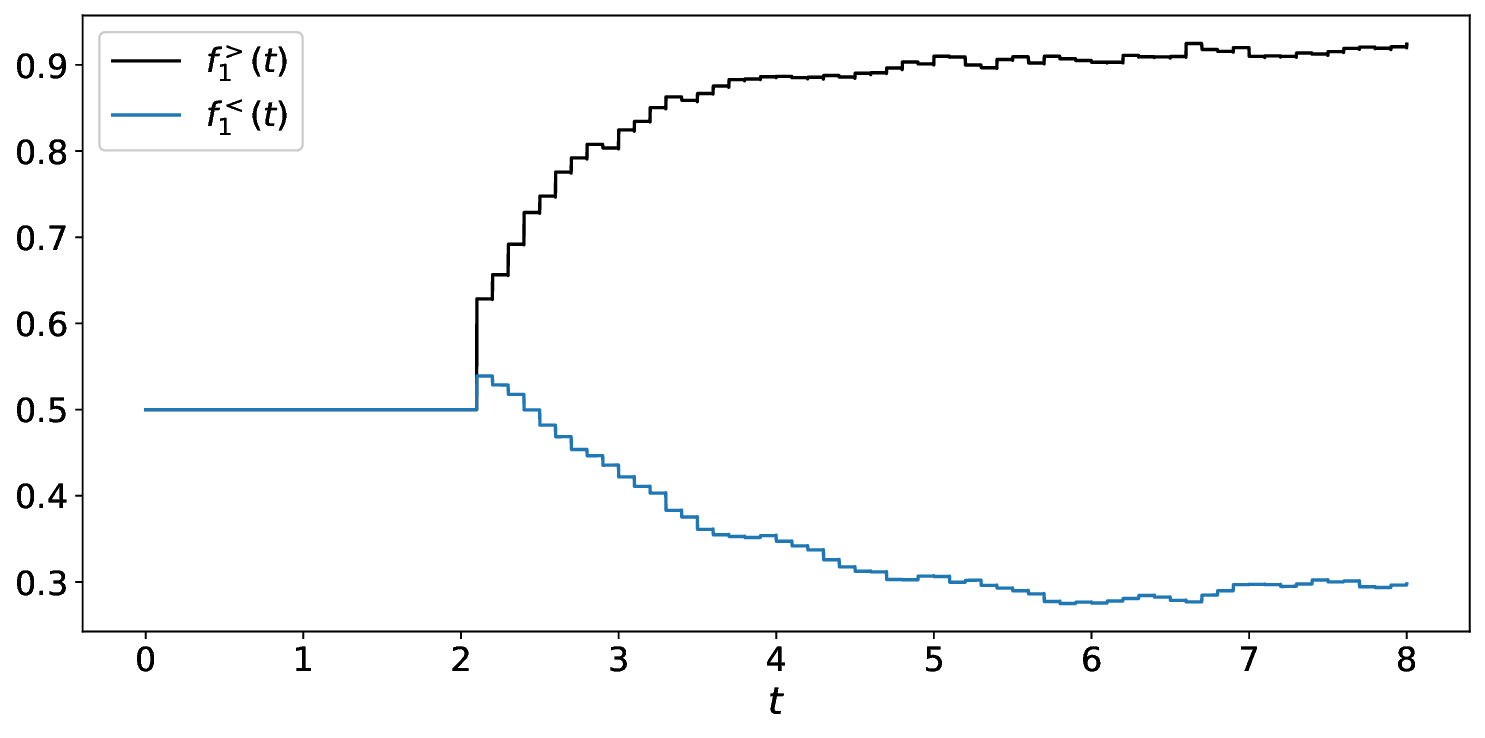} 
  \caption{Effective share of users on platform $1$ in the case where $\beta^{(1)}=2, \beta^{(2)}=1$. $\theta=0.1$, so that approximately 90\% of individuals have $a_i<\theta$. }
  \label{fig:seg_grandi_piccoli}
\end{figure}

To understand the benefits of the multi-platform reported in the main text, we define the following quantities.
\begin{equation}
\begin{aligned}
    f_\gamma^>(t)&=\frac{\sum_{i=1}^N \rho_i^{(\gamma)}(t) \mathcal{H}(a_i-\theta)}{\sum_{i=1}^N \mathcal{H}(a_i-\theta)} \\
    f_\gamma^<(t)&=\frac{\sum_{i=1}^N \rho_i^{(\gamma)}(t) \mathcal{H}(\theta-a_i)}{\sum_{i=1}^N \mathcal{H}(\theta-a_i)},
\end{aligned}
\end{equation}
where $\mathcal{H}$ is the Heaviside function. In words, $f_\gamma^>(t)$ (resp. $f_\gamma^<(t)$) is the ``effective'' share of users on platform $\gamma$ whose activity is above (resp. below) a threshold $\theta$. 
The idea is to understand whether these two classes of users exhibit qualitatively different behavior, notwithstanding their a-priori equal preferences (i.e. $\phi_i=\phi$ and $r_i=r$ for all users).
Let us consider the 2 platform case reported in the main text, where $\beta^{(1)}=2, \beta^{(2)}=1$. In Fig.~\ref{fig:seg_grandi_piccoli}, we plot $f_1^>(t)$ and $f_1^<(t)$ for $\theta=0.1$, which roughly divide the activity profile in 90 \% below the threshold and 10\% above. Highly active users deterministically prefer the more homophilic platform ($f_1^> \approx 1$), while less active users tend to explore both, though they prevalently occupy the platform with more cross-cutting content ($f_1^< \approx 0.3$)

This is consistent with what we observe in reality, i.e. people with extreme opinions tend to have a more restricted outlet of like-minded sources, which further polarizes their view. On the other hand, more moderate users typically exhibit a more diverse news diet.

\section{Robustness with respect to desired distinctiveness}
\begin{figure}[ht!]
    \centering

    \subfigure[]{
        \includegraphics[width=0.5\linewidth]{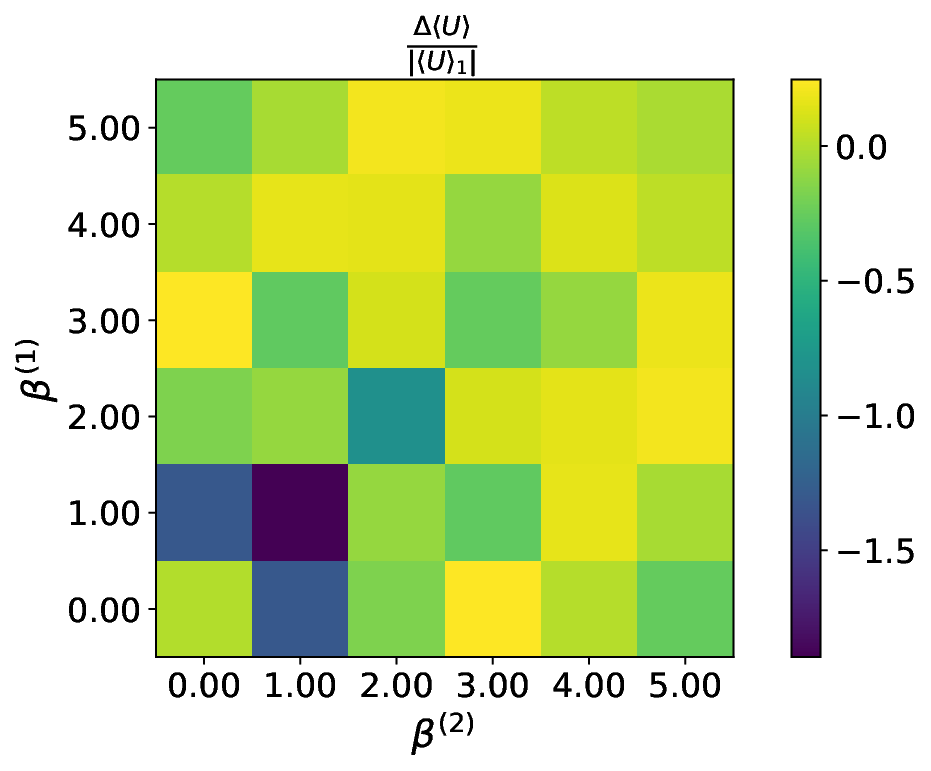}
        \label{fig:u2plat2}
    }
    \hfill
    \subfigure[]{
        \includegraphics[width=0.5\linewidth]{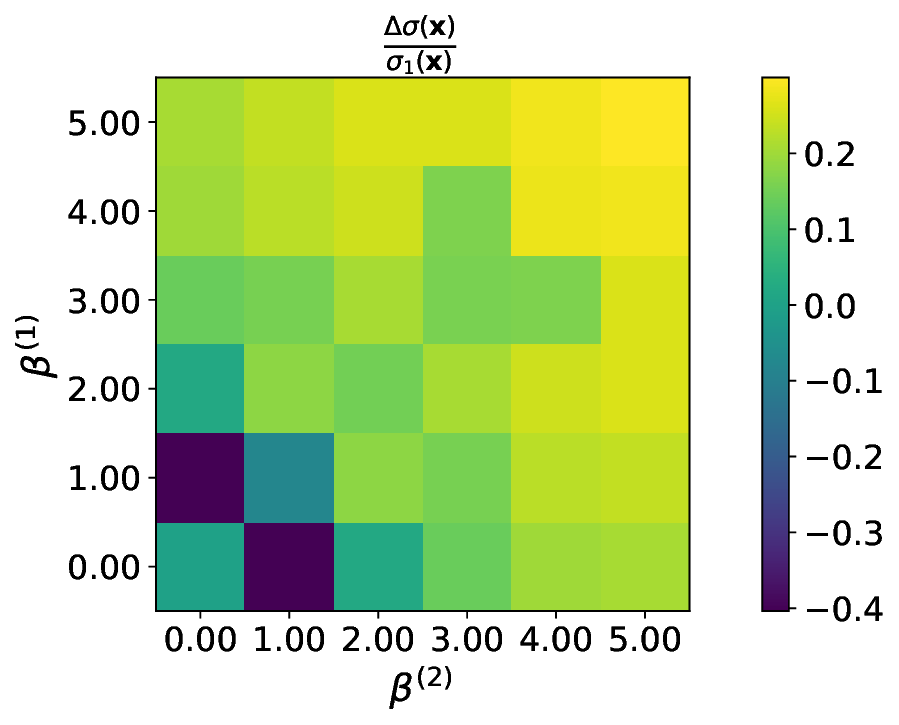}
        \label{fig:std2plat2}
    }
        \caption{Relative increment of (a) satisfaction and (b) opinions' standard deviation, when passing from $\Gamma=1$ to $\Gamma=2$ platforms, for $\phi=0.1$. In this regime, high values of both $\beta^{(1)}$ and $\beta^{(2)}$ lead to an increased users' satisfaction but also in an increased polarization (i.e. standard deviation).}
    \label{fig:rhodin2plat2}
\end{figure}

Here, we want to show how satisfaction and opinions standard deviation change with respect to the single-platform case for $\phi=0.1$. The main physical difference with respect to the case presented in the main text (where $\phi=0.2$) is that here users have an ``halved'' desire for differentiation, meaning that their maximal satisfaction requires a much high number of interactions contributing to assimilation, with respect to those contributing to differentiation. Figure~\ref{fig:rhodin2plat2} shows indeed how users' utility is maximized in those regimes for which they are more exposed to like-minded peers, i.e. for high values of $\beta^{(1)}$ and $\beta^{(2)}$. However, in this case, maximizing users' satisfaction leads to an increase in polarization, meaning that if users desire high confirmation opinions tend to extremize.

\end{appendices}

\end{document}